\documentstyle[preprint,aps]{revtex}
\tighten
\draft
\begin{document}
\widetext
\input epsf
\preprint{CERN-TH/97-170, HUTP-97/A036, NUB 3164}
\bigskip
\bigskip
\title{A Remark on Dilaton Stabilization}
\medskip
\author{Gia Dvali$^{1}$\footnote{E-mail: 
georgi.dvali@cern.ch} and 
Zurab Kakushadze$^{2,3}$\footnote{E-mail: 
zurab@string.harvard.edu}}

\bigskip
\address{$^1$Theory Division, CERN, CH-1211, Geneva 22, Switzerland\\
$^2$Lyman Laboratory of Physics, Harvard University, Cambridge, 
MA 02138\\
$^3$Department of Physics, Northeastern University, Boston, MA 02115}
\date{\today}
\bigskip
\medskip
\maketitle

\begin{abstract}

{}Dilaton stabilization may occur in a theory based on a single asymptotically free gauge group with matter due to an interplay between quantum modification of the moduli space and tree-level superpotential. We present a toy model where such a mechanism is realized. Dilaton stabilization in this mechanism tends to occur at strong coupling values unless some unnatural adjustment of parameters is involved.

\end{abstract}
\pacs{}
\narrowtext

{}The gauge and gravitational couplings in the effective field theory of any string derived model are determined by vevs of certain moduli. Thus, for example, in perturbative heterotic superstring the gauge $g_a$ and gravitational (that is, string) $g$ couplings at the string scale are related to each other via $K_a g^2_a=g^2$, where $K_a$ are the current algebra levels of the corresponding gauge subgroups $G_a$. The string coupling is determined by the vev of the dilaton field $S$: $\langle S\rangle=1/g^2+i\theta/8\pi^2$ (where $\theta$ is the vacuum angle). Perturbatively $S$ is a modulus field, and its expectation value is undetermined. The dilaton stabilization must therefore have non-perturbative origin.

{}One possibility is to consider the standard ``race-track'' scenario \cite{kras}, where non-perturbative superpotential (which is exponential in $S$) is generated by gaugino condensation. Dilaton stabilization then requires presence of at least two 
gauge groups giving rise to different exponentials\footnote{In Ref \cite{que} exponential  contributions to the superpotential were argued to also arise in {\em non}-asymptotically-free gauge theories.} in the superpotential\footnote{This mechanism requires rather large gauge groups to achieve weak coupling stabilization. Such large gauge groups can {\em a priori} appear in non-perturbative string vacua. This idea was recently discussed in the context of $F$-theory in Ref \cite{KL}.}. 

{}In this note we argue that dilaton stabilization may occur in a theory based on a {\em single} asymptotically free gauge group with matter\footnote{Dilaton stabilization might be possible in a theory with a single gaugino condensate \cite{BD} if the K{\"a}hler potential receives large non-perturbative corrections.}. As we discuss below, the dilaton is stabilized here due to an interplay between {\em quantum modification} of the moduli space and {\em tree-level} couplings of the gauge invariants with additional gauge singlets. Here we present a toy model where such a mechanism is realized. 

{}Thus, consider a theory with $SU(N)$ gauge group and $N$ flavors $Q^i,{\tilde Q}_{\bar j}$ ($i,{\bar j}=1,\dots,N$). The gauge invariant degrees of freedom are mesons $M^i_{\bar j}\equiv Q^i {\tilde Q}_{\bar j}$, and baryons $B\equiv\epsilon_{{i_1}\dots {i_{N_c}}}Q^{i_1}\cdots Q^{i_{N_c}}$ and ${\tilde B}\equiv\epsilon^{{{\bar j}_1}\dots {{\bar j}_{N_c}}} 
{\tilde Q}_{{\bar j}_1}\cdots {\tilde Q}_{{\bar j}_{N_c}}$. The classical moduli space in this theory receives quantum corrections which can be accounted for via the following superpotential \cite{Seiberg}
\begin{eqnarray}\label{non-pert}
 {\cal W}_{non-pert}=A\left(\det(M)-
 B{\tilde B}-\Lambda^{2N}\right)~,
\end{eqnarray}
where $A$ is the Lagrange multiplier ($A\Lambda^{2N}=W_aW_a$ is the ``glue-ball'' field), and $\Lambda\equiv\exp(-4\pi^2 S/N)$ is the dynamically generated scale of the theory. (Here for simplicity we take the $SU(N)$ current algebra level to be 1.) The quantum constraint then follows from the $F$-flatness condition for the field $A$ and reads:
\begin{eqnarray}\label{quantum}
 \det(M)-B{\tilde B}-\Lambda^{2N}=0~.
\end{eqnarray}
Note that with just this constraint the dilaton is not stabilized. If, however, $\det(M)$, $B$ and ${\tilde B}$ are fixed via some other dynamics, then the quantum constraint (\ref{quantum}) will fix the dilaton vev (provided that $0<\vert\det(M)-B{\tilde B}\vert<1$).

{}The simplest possibility here is to require that there be present tree-level contributions to the superpotential (which could be both renormalizable and non-renormalizable couplings). Note that {\em a priori} they need not even respect any of the global symmetries of the above quantum moduli space. Upon inclusion of such couplings into the superpotential, dilaton may be stabilized (without breaking supersymmetry).

{}As a simple toy example consider the following tree-level superpotential:
\begin{equation}\label{tree}
{\cal W}_{tree} = YB+{\tilde Y}{\tilde B}+(\lambda-X)\det(M) + {\rho \over {n+1}}X^{n+1}~,
\end{equation}
where $X,Y,{\tilde Y}$ are additional chiral superfields (which are singlets of $SU(N)$), and $\lambda,\rho$ are some couplings. The superpotential is given by ${\cal W}={\cal W}_{non-pert}+{\cal W}_{tree}$. The $F$-flatness conditions for the singlets $Y,{\tilde Y}$ imply that $B={\tilde B}=0$. The $F$-flatness condition for the singlet $X$ implies that  
$\det(M)=\rho X^n$. Note that if $X=0$ then the quantum constraint (\ref{quantum}) cannot be satisfied for any finite values of the dilaton vev $S$. Thus, $X=0$ lies on a non-supersymmetric {\em runaway} branch. There is, however, a family of supersymmetric vacua in the moduli space. First note that the dilaton $F$-flatness condition implies $A=0$.  Next, if $X\not=0$, then it follows that $\det(M)\not=0$. On the other hand, the $F$-flatness conditions for the mesons $M^i_{\bar j}$ imply that $(\lambda-X){\cal M}_i^{\bar j}=0$, where ${\cal M}_i^{\bar j}\equiv \partial\det(M)/\partial M^i_{\bar j}$. This implies that ${\cal M}_i^{\bar j}\equiv 0$ unless $X=\lambda$. Since $\det(M)=M^i_{\bar j} {\cal M}_i^{\bar j}/N$, it follows from the above $F$-flatness conditions that if $X\not=0$ then $X=\lambda$. Finally, the quantum constraint (\ref{quantum}) along with the rest of the $F$-flatness conditions we have just discussed implies that $\Lambda^{2N}=\det(M)=
\rho X^n=\rho\lambda^n$ provided that $X=\lambda$. 
Thus, the above superpotential has a family of supersymmetric vacua with
\begin{equation}\label{vacuum}
 A=B={\tilde B}=0~,~~~X=\lambda~,~~~
 S={n\over 8\pi^2}\log(\tau)~,~~~\det(M)={1\over \tau^n}
\end{equation}      
provided that $\vert\tau\vert>1$. Here $1/\tau\equiv\rho^{1/n}\lambda$. Note that this family of supersymmetric vacua is parametrized by the meson vevs $M^i_{\bar j}$ subject to the constraint $\det(M)=1/\tau^n$. Thus, there are $N^2-1$ left-over flat directions. The other vevs, including the dilaton, however, are fixed.  Note that this family of supersymmetric vacua is separated from the runaway branches by potential barriers.

{}We see that in this example the dilaton is stabilized at strong coupling values unless $n\sim 8\pi^2$ (assuming that $\log(\vert\tau\vert)\sim1$), which looks unnatural. One may attempt to find an ``improvement'' for the above toy model (at the expense of introducing additional singlet fields and tree-level couplings) which would allow weak coupling stabilization. However, all the models we have constructed so far look rather contrived. There appears to be a simple reason for this. The entire idea of dilaton stabilization described in this note is based on the quantum constraint (\ref{quantum}). To stabilize the dilaton one requires that the quantity ${\cal C}\equiv\det(M)-B{\tilde B}$ is stabilized at a non-zero value via some additional dynamics. The dilaton enters Eq (\ref{quantum}) in the combination $\Lambda^{2N}=\exp(-8\pi^2 S)$, and the stabilized value of $S$ is given by $S=-{1\over 8\pi^2} \log({\cal C})$. Unless ${\cal C}$ is an exponentially small number, the stabilized value of $S$ will always be at strong coupling.
 
{}The above toy model defined by Eq (\ref{tree}) is not generic. However, generic models
can also be constructed. For instance, consider the following tree-level superpotential:
\begin{equation}
 {\cal W}_{tree}=Xf(\det(M),B{\tilde B})+Yg(\det(M),B{\tilde B})~,
\end{equation}
where $X,Y$ are singlet superfields both with $R$-charge 2, and $f,g$ are arbitrary polynomials of their arguments $\det(M)$ and $B{\tilde B}$. This superpotential respects all the symmetries 
of Eq (\ref{non-pert}). The dilaton is stabilized without breaking supersymmetry provided that the equation $f=g=0$ has isolated solutions with $0<\vert\det(M)-B{\tilde B}\vert<1$.

\acknowledgments

{}We would like to thank Luis {\'A}lvarez-Gaum{\'e}, Ignatios Antoniadis, Ram Brustein, Savas Dimopoulos, Emilian Dudas, Andrei Johansen, Vadim Kaplunovsky, Alex Pomarol, Lisa Randall, Riccardo Rattazzi, Tom Taylor, and Henry Tye for discussions. The work of Z.K. was supported in part by the grant NSF PHY-96-02074, and the DOE 1994 OJI award. Z.K. would like to thank CERN Theory Division for their kind hospitality while parts of this work were completed. Z.K. would also like to thank Albert and Ribena Yu for financial support.

\end{document}